\documentclass[a4paper]{jpconf}
\usepackage{graphicx}
\begin{document}
\title{Nonlinear statistical effects in relativistic mean field theory}

\author{G Gervino$^{1,3}$, A Lavagno$^{2,3}$, D Pigato$^{2,3}$}

\address{
$^1$Dipartimento di Fisica, Universit\`a di
Torino, I-10126 Torino, Italy \\
$^2$Dipartimento di Fisica, Politecnico di Torino, C.so
Duca degli Abruzzi 24,  Italy \\
$^3$Istituto Nazionale di Fisica Nucleare (INFN), Sezione di
Torino, Italy}
%\ead{andrea.lavagno@polito.it}

\begin{abstract}
We investigate the relativistic mean field theory of nuclear
matter at finite temperature and baryon density taking into
account of nonlinear statistical effects, characterized by
power-law quantum distributions. The analysis is performed by
requiring the Gibbs conditions on the global conservation of
baryon number and electric charge fraction.
We show that such nonlinear statistical effects play a crucial role in
the equation of state and in the formation of mixed phase also for
small deviations from the standard Boltzmann-Gibbs statistics.
\end{abstract}

\section{Introduction}
Several experimental observations and theoretical calculations clearly indicate that
hadrons dissociate into a plasma of their elementary constituents,
quarks and gluons (QGP), at density several times the nuclear
matter density and/or at temperature above few hundreds MeV. Such a QGP is
expected to have occurred in the early stages of the Universe and
can be found in dense and hot stars, neutron stars,
nucleus-nucleus high energy collisions where heavy ions are
accelerated to relativistic energies~\cite{hwa}. After collision,
a fireball is created which may realize the conditions of the QGP.
The plasma then expands, cools, freezes-out into hadrons, photons,
leptons that are detected and analyzed \cite{biro08}.

It is a rather common opinion that, because of the extreme
conditions of density and temperature in ultrarelativistic heavy
ion collisions, memory effects and long--range color interactions
give rise to the presence of non--Markovian processes in the
kinetic equation affecting the thermalization process toward
equilibrium as well as the standard equilibrium distribution
\cite{hei,biro,albe}. A rigorous determination of the conditions
that produce a nonextensive behavior, due to memory effects and/or
long--range interactions,  should be based on microscopic
calculations relative to the parton plasma originated during the
high energy collisions. At this stage we limit ourselves to
consider the problem from a qualitative point of view on the basis
of the existing theoretical calculations and experimental
evidences.

On the other hand, over the last years, there has been an
increasing evidence that the generalized non-extensive statistical
mechanics, proposed by Tsallis \cite{tsallis,GMTsallis,book2,kodama}
and characterized by a power-law stationary particle distribution,
can be considered as a
 basis for a theoretical framework appropriate to incorporate, at
least to some extent and without going into microscopic dynamical
description, long-range interactions, long-range microscopic
memories and/or fractal space-time constraints. A considerable
variety of physical issues show a quantitative agreement between
experimental data and theoretical analysis based on Tsallis'
thermostatistics. In particular, there is a growing interest in
high energy physics applications of non-extensive statistics
\cite{bediaga,beck,rafelski,wilk1,plb2001,pla2002,biroprl05}. Several
authors outline the possibility that experimental observations in
relativistic heavy-ion collisions can reflect non-extensive
statistical mechanics effects during the early stage of the
collisions and the thermalization evolution of the system
\cite{albe1,biro04,wilk2,physicaA2008,cley}.
In this context, it is relevant to observe that the statistical
origin of the nonextensive statistics lies in the deformation of
the Boltzmann entropy.
From the above considerations, it appears reasonable that in
regime of high density and temperature both hadron and quark-gluon
Equation of State (EOS) can be sensibly affected by nonextensive
statistical effects \cite{physicaEOS,wilknjl}. Furthermore, in this
context it is very remarkable to observe that the relevance of
these effects on the relativistic hadronic equation of state has
also been recently investigated in Ref.~\cite{silva}.

The aim of this paper is to study the behavior of the nuclear
equation of state at finite temperature and baryon density and to
explore the existence of a hadron-quark mixed phase at a fixed
value of the proton fraction $Z/A$.

\section{Nonextensive hadronic and quark-gluon equation of state}
In this Section we study the nonextensive hadronic EOS in the
framework of a relativistic mean field theory in which nucleons
interact through the nuclear force mediated by the exchange of
virtual isoscalar-scalar ($\sigma$), isoscalar-vector ($\omega$)
and isovector-vector ($\rho$) meson fields
\cite{walecka,boguta,glen}.

The nonlinear Lagrangian density describing hadronic matter can be written
as
\begin{eqnarray}
{\cal L}={\cal L}_{QHD}+{\cal L}_{\rm qfm} \, ,\label{totl}
\end{eqnarray}
where \cite{glen}
\begin{eqnarray}\label{eq:1}
\!\!\!\!\!\!\!\!\!\!\!\!\!\!\!{\cal L}_{QHD}&=&
\bar{\psi}[i\gamma_{\mu}\partial^{\mu}-(M- g_{\sigma}\sigma)
-g{_\omega}\gamma_\mu\omega^{\mu}-g_\rho\gamma^{\mu}\vec\tau\cdot
\vec{\rho}_{\mu}]\psi
+\frac{1}{2}(\partial_{\mu}\sigma\partial^{\mu}\sigma-m_{\sigma}^2\sigma^2)
\nonumber\\
\!\!\!\!\!\!\!\!\!\!\!\!\!\!\!&&-U(\sigma)+\frac{1}{2}m^2_{\omega}\omega_{\mu}
\omega^{\mu}
+\frac{1}{2}m^2_{\rho}\vec{\rho}_{\mu}\cdot\vec{\rho}^{\;\mu}
-\frac{1}{4}F_{\mu\nu}F^{\mu\nu}
-\frac{1}{4}\vec{G}_{\mu\nu}\vec{G}^{\mu\nu}\,,
\end{eqnarray}
and $M=939$ MeV is the vacuum baryon mass. The field strength
tensors for the vector mesons are given by the usual expressions
$F_{\mu\nu}\equiv\partial_{\mu}\omega_{\nu}-\partial_{\nu}\omega_{\mu}$,
$\vec{G}_{\mu\nu}\equiv\partial_{\mu}\vec{\rho}_{\nu}-\partial_{\nu}\vec{\rho}_{\mu}$,
and $U(\sigma)$ is a nonlinear potential of $\sigma$ meson
\begin{eqnarray}
U(\sigma)=\frac{1}{3}a\sigma^{3}+\frac{1}{4}b\sigma^{4}\,,
\end{eqnarray}
usually introduced to achieve a reasonable compression modulus for
equilibrium nuclear matter.

Following Ref.s \cite{muller_npa,lava_prc}, ${\cal L}_{\rm qfm}$ in
Eq.(\ref{totl}) is related to a (quasi) free gas of pions with an
effective chemical potential (see below for details).

The field equations in a mean field approximation are
\begin{eqnarray}
&&(i\gamma_{\mu}\partial^{\mu}-(M- g_{\sigma}\sigma)-
g_\omega\gamma^{0}\omega-g_\rho\gamma^{0}{\tau_3}\rho)\psi=0\,, \\
&&m_{\sigma}^2\sigma+ a{{\sigma}^2}+ b{{\sigma}^3}=
g_\sigma<\bar\psi\psi>=g_\sigma{\rho}_S\,, \\
&&m^2_{\omega}\omega=g_\omega<\bar\psi{\gamma^0}\psi>=g_\omega\rho_B\,,\\
&&m^2_{\rho}\rho=g_\rho<\bar\psi{\gamma^0}\tau_3\psi>=g_\rho\rho_I\,,
\label{eq:MFT}
\end{eqnarray}
where $\sigma=\langle\sigma\rangle$,
$\omega=\langle\omega^0\rangle$ and $\rho=\langle\rho^0_3\rangle$
are the nonvanishing expectation values of meson fields, $\rho_I$
is the total isospin density, $\rho_B$ and $\rho_S$ are the baryon
density and the baryon scalar density, respectively. They are
given by
\begin{eqnarray}
&&\rho_{B}=2 \sum_{i=n,p} \int\frac{{\rm
d}^3k}{(2\pi)^3}[n_i(k)-\overline{n}_i(k)]\,, \label{eq:rhob} \\
&&\rho_S=2 \sum_{i=n,p} \int\frac{{\rm
d}^3k}{(2\pi)^3}\,\frac{M_i^*}{E_i^*}\,
[n_i^q(k)+\overline{n}_i^{\,q}(k)]\,, \label{eq:rhos}
\end{eqnarray}
where $n_i(k)$ and $\overline{n}_i(k)$ are the $q$-deformed
fermion particle and antiparticle distributions:
\begin{eqnarray}
n_i(k)=\frac{1} { [1+(q-1)\,\beta(E_i^*(k)-\mu_i^*)
]^{1/(q-1)} + 1} \label{eq:distribuz} \, , \\
\overline{n}_i(k)=\frac{1} {[1+(q-1)\,\beta(E_i^*(k)+\mu_i^*)
]^{1/(q-1)} + 1} \, . \label{eq:distribuz2}
\end{eqnarray}

The nucleon effective energy is defined as
${E_i}^*(k)=\sqrt{k^2+{{M_i}^*}^2}$, where ${M_i}^*=M_{i}-g_\sigma
\sigma$. The effective chemical potentials $\mu_i^*$  are given in
terms of the meson fields as follows
\begin{eqnarray}
\mu_i^*={\mu_i}-g_\omega\omega -\tau_{3i} g_{\rho}\rho \, ,
\label{mueff}
\end{eqnarray}
where $\mu_i$ are the thermodynamical chemical potentials
$\mu_i=\partial\epsilon/\partial\rho_i$. At zero temperature they
reduce to the Fermi energies $E_{Fi} \equiv
\sqrt{k_{Fi}^2+{M_i^*}^2}$ and the nonextensive statistical effects
disappear. The meson fields are obtained as a solution of the field
equations in mean field approximation and the related meson-nucleon
couplings ($g_\sigma$, $g_\omega$ and $g_\rho$) are the free
parameters of the model. In the following, they will be fixed to the
parameters set marked as GM2 of Ref. \cite{glen}.

The thermodynamical quantities can be obtained from the
thermodynamic po\-tential in the standard way. More explicitly,
the baryon pressure $P_B$ and the energy density $\epsilon_B$ can
be written as
\begin{eqnarray}
\hspace{-2cm}&&P_B=\frac{2}{3} \sum_{i=n,p} \int \frac{{\rm
d}^3k}{(2\pi)^3} \frac{k^2}{E_{i}^*(k)}
[n_i^q(k)+\overline{n}_i^q(k)] -\frac{1}{2}m_\sigma^2\sigma^2 -
U(\sigma)+
\frac{1}{2}m_\omega^2\omega^2+\frac{1}{2}m_{\rho}^2 \rho^2\,,\label{eq:eos}\\
\hspace{-2cm}&&\epsilon_B= 2 \sum_{i=n,p}\int \frac{{\rm
d}^3k}{(2\pi)^3}E_{i}^*(k) [n_i^q(k)+\overline{n}_i^q(k)]
+\frac{1}{2}m_\sigma^2\sigma^2+U(\sigma)
+\frac{1}{2}m_\omega^2\omega^2+\frac{1}{2}m_{\rho}^2 \rho^2\, .
\label{eq:eos2}
\end{eqnarray}

It is important to observe that Eq.s(\ref{eq:rhos}),
(\ref{eq:eos}) and (\ref{eq:eos2}) apply to $n_i^q\equiv(n_i)^q$
rather than $n_i$ itself, this is a direct consequence of the
basic prescription related to the $q$-mean expectation value in
nonextensive statistics \cite{book2,pla2002}. In addition, since all equations
must be solved in a self-consistent way, the presence of
nonextensive statistical effects influences the many-body
interaction mediated by the meson fields.

Especially in regime of low density and high temperature the
contribution of the lightest mesons to the thermodynamical
potential (and, consequently, to the other thermodynamical
quantities) becomes relevant. As quoted before, we have included the contribution of pions
considering them as a (quasi) ideal gas of nonextensive bosons
with effective chemical potentials expressed in terms of the
corresponding effective baryon chemical potentials \cite{lava_prc}.

\begin{figure}
\begin{center}
\resizebox{1.0\textwidth}{!}{%
\includegraphics{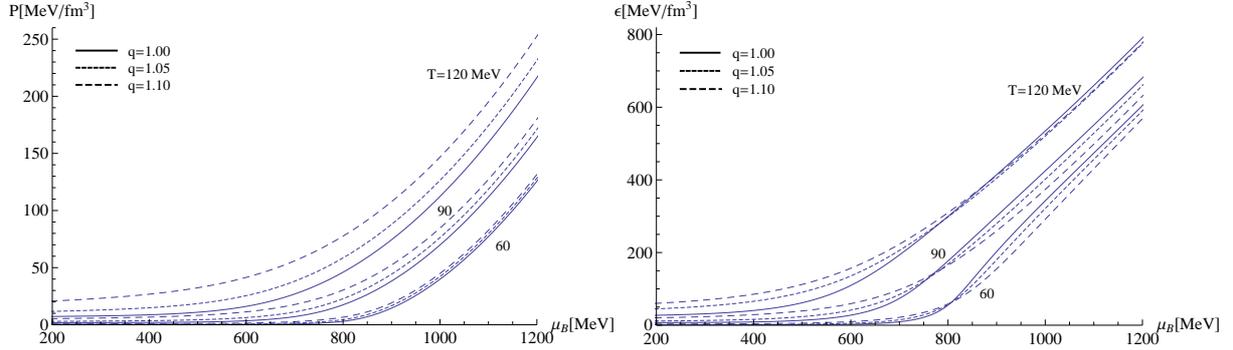}
} \caption{Pressure (left panel) and energy density (right panel)
versus baryon chemical potential for different values of
temperature and $q$.} \label{fig:P} \end{center}
\end{figure}

In Fig. \ref{fig:P}, the total pressure $P$ and energy density
$\epsilon$ are plotted as a function of $\mu_B$ for different
values of temperature and $q$. The different behavior from $P$ and
$\epsilon$ reflects essentially the nonlinear combinations of the
meson fields and the different functions under integration in Eq.s
(\ref{eq:eos}) and (\ref{eq:eos2}). Concerning the pressure, we
have that becomes stiffer by increasing the $q$ parameter. On the
other hand, the behavior of the energy density presents features
very similar to the $\sigma$ field one. At low $\mu_{B}$,
nonextensive effects make the energy density greater with respect
to the standard case. At medium-high $\mu_B$, the standard ($q=1$)
component of the energy density becomes dominant, this effect is
essentially due to the reduction of the $\sigma$ field for $q>1$.
The intersection point depends, naturally, on the physical
parameters of the system.

Concerning the quark-gluon EOS, we use the MIT bag model
\cite{mit}. In this model, quark matter is described as a gas of
free quarks with massless up and down quarks. All the
non-perturbative effects are simulated by the bag constant $B$
which represents the pressure of the vacuum. Following this line,
the pressure, energy density and baryon number density for a
relativistic Fermi gas of quarks in the framework of nonextensive
statistics can be written, respectively, as
\begin{eqnarray}
&P_q& =\frac{\gamma_f}{3} \sum_{f=u,d} \int^\infty_0 \frac{{\rm
d}^3k}{(2\pi)^3} \,\frac{k^2}{e_f}\,
[n_f^q(k)+\overline{n}_f^q(k)]
%\nonumber \\&&\;\;\;\;
-B\,, \label{bag-pressure}\\
&\epsilon_q& =\gamma_f \sum_{f=u,d}  \int^\infty_0 \frac{{\rm
d}^3k}{(2\pi)^3} \,e_f\, [n_f^q(k)+\overline{n}_f^q(k)]
\label{bag-energy}
%\nonumber \\&&\;\;\;\;
+B\,, \\
&\rho_q& =\frac{\gamma_f}{3} \sum_{f=u,d} \int^\infty_0 \frac{{\rm
d}^3k}{(2\pi)^3}  \,[n_f(k)-\overline{n}_f(k)]\, ,
\label{bag-density}
\end{eqnarray}
where the quark degeneracy for each flavor is $\gamma_f=6$,
$e_f=(k^2+m_f^2)^{1/2}$, $n_f(k)$ and $\overline{n}_f(k)$ are the
$q$-deformed particle and antiparticle quark distributions
\begin{eqnarray}
n_f(k)=\frac{1} { [1+(q-1)(e_f(k)-\mu_f)/T
]^{1/(q-1)} + 1} \, , \\
\overline{n}_f(k)=\frac{1}{[1+(q-1)(e_f(k)+\mu_f)/T ]^{1/(q-1)} +
1} \, .
\end{eqnarray}

Similar expressions for the pressure and the energy density can be
written for gluons treating them as a massless $q$-deformed Bose
gas with zero chemical potential. Explicitly, we can calculate the
nonextensive pressure $P_g$ and energy density $\epsilon_g$ for
gluons as
\begin{eqnarray}
&P_g& =\frac{\gamma_g}{3} \int^\infty_0 \frac{{\rm
d}^3k}{(2\pi)^3}
\,\frac{k}{[1+(q-1)\,k/T]^{q/(q-1)} - 1}\,, \label{gluon-press}\\
&\epsilon_g& =3\, P_g \, , \label{gluon-energy}
\end{eqnarray}
with the gluon degeneracy factor $\gamma_g=16$. In the limit
$q\rightarrow 1$, one recovers the usual analytical expression:
$P_g=8\pi^2/45\,T^4$.

Let us note that, since one has to employ the fermion (boson)
nonexten\-sive distributions, the results are not analytical, even
in the massless quark approximation. Hence a numerical evaluations
of the integrals in Eq.s~(\ref{bag-pressure})--(\ref{bag-density})
and (\ref{gluon-press}) must be performed.

\begin{figure}
\begin{center}
\resizebox{0.6\textwidth}{!}{%
  \includegraphics{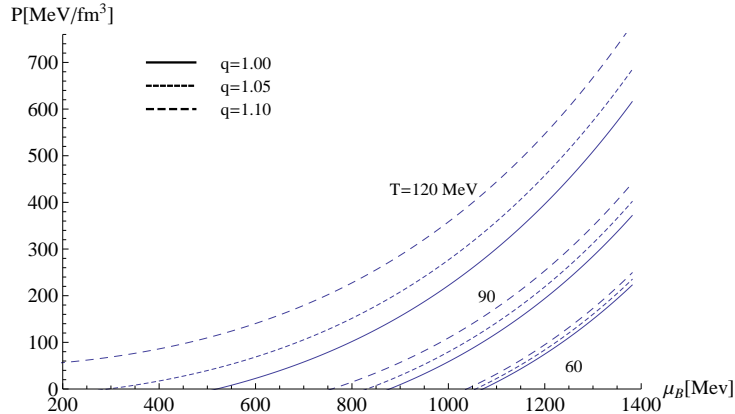}
} \caption{Pressure of the quark-gluon phase as a function of
baryon chemical potential for different values of temperature and
$q$.} \label{fig:PQGP}
\end{center}
\end{figure}

In Fig. \ref{fig:PQGP}, we report the total pressure as a function
of the baryon chemical potential for massless quarks and gluons,
for different values of $q$ and at fixed value of $Z/A=0.4$. The
bag constant is set equal to $B^{1/4}$=190 MeV. In presence of
nonextensive effects, as in the case of hadronic phase, the
pressure is significantly increased even for small deviations from
standard statistics.

\section{The hadron to quark-gluon phase transition}\label{mp}

In this Section we investigate the hadron-quark phase transition
at finite temperature and baryon chemical potential by means of the
previous relativistic EOSs. Lattice calculations predict a critical
phase transition temperature $T_c$ of about 170 MeV, corresponding
to a critical energy density $\epsilon_c\approx$ 1 GeV/fm$^3$
\cite{hwa}. In a theory with only gluons and no quarks, the
transition turns out to be of first order. In nature, since the $u$
and $d$ quarks have a small mass, while the strange quark has a
somewhat larger mass, the phase transition is predicted to be a
smooth cross over. However, since it occurs over a very narrow range
of temperatures, the transition, for several practical purposes, can
still be considered of first order. Indeed the lattice data with 2
or 3 dynamical flavours are not precise enough to unambigously
control the difference between the two situations. Thus, by
considering the deconfinement transition at finite density as a the
first order one, a mixed phase can be formed, which is typically
described using the two separate equations of state, one for the
hadronic and one for the quark phase.

The phase transition is described by using the Gibbs formalism applied to systems where
more than one conserved charge is present \cite{glenprd}. In fact,
because we are going to describe the nuclear EOS, we have to
require the global conservation of two "charges": baryon number
and electric charge. Each conserved charge has a conjugated
chemical potential and the systems is described by two independent
chemical potentials: $\mu_B$ and $\mu_C$. The structure of the
mixed phase is obtained by imposing the following Gibbs conditions
for chemical potentials and pressure \cite{dlplb,ditoro,prl}
\begin{eqnarray}
&&\mu_B^{(H)} = \mu_B^{(Q)} \, , \ \ \  \mu_C^{(H)} = \mu_C^{(Q)}
\, , \\
&&P^H (T,\mu_B,\mu_C)=P^Q (T,\mu_B,\mu_C) \, .
\end{eqnarray}
Therefore, at a given baryon density $\rho_B$ and at a given net
electric charge density $\rho_C=Z/A\, \rho_B$, the chemical
potentials $\mu_B$ are $\mu_C$ are univocally determined by the
following equations
%
%\begin{eqnarray}
%&&\rho_B=(1-\chi)\,\sum_{i=p,n} b_i\,\rho_i^H(T,\mu_B,\mu_C)
%+\chi \,\sum_{i=u,d} b_i\,\rho_i^Q(T,\mu_B,\mu_C) \, ,\\
%&&\rho_C=(1-\chi)\,\sum_{i=p,n} c_i\, \rho_i^H(T,\mu_B,\mu_C)
%+\chi \,\sum_{i=u,d} c_i\,\rho_i^Q(T,\mu_B,\mu_C) \, ,
%\end{eqnarray}
\begin{eqnarray}
&&\rho_B=(1-\chi)\,\rho_B^H(T,\mu_B,\mu_C)
+\chi \,\rho_B^Q(T,\mu_B,\mu_C) \, ,\\
&&\rho_C=(1-\chi)\,\rho_C^H(T,\mu_B,\mu_C) +\chi
\,\rho_C^Q(T,\mu_B,\mu_C) \, ,
\end{eqnarray}
where $\rho_B^{H(Q)}$ and $\rho_C^{H(Q)}$ are, respectively, the
net baryon and electric charge densities in the hadronic (H) and
in the quark (Q) phase and $\chi$ is the fraction volume of
quark-gluon matter in the mixed phase. In this way we can find out
the phase coexistence region, for example, in the $(T,\mu_B)$
plane. We are particularly interested in the lower baryon density
(baryon chemical potential) border, i.e. the first critical
transition density $\rho_{\rm cr}^I$ ($\mu_{\rm cr}^I$), in order
to check the possibility of reaching such conditions in a
transient state during a heavy-ion collision at relativistic
energies.

\begin{figure}
\begin{center}
\resizebox{1.0\textwidth}{!}{%
  \includegraphics{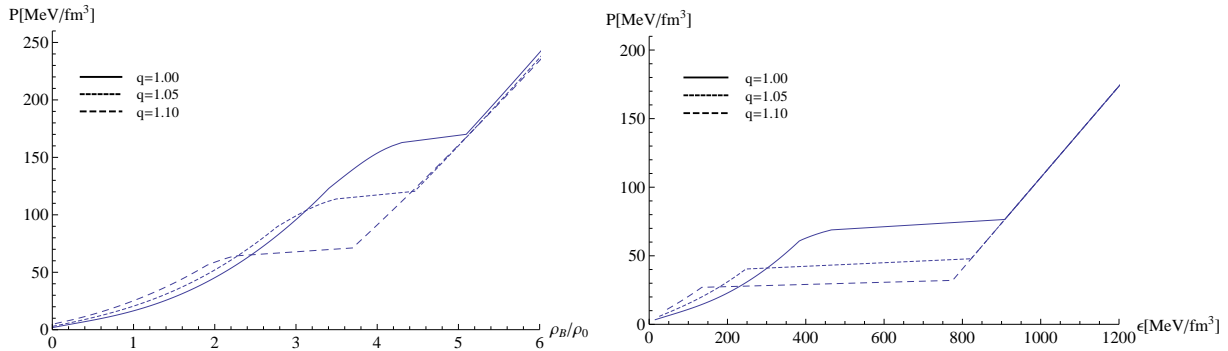} } \caption{Pressure as a function of baryon density
  (left panel) and energy density (right panel) in the mixed
  phase for different values of $q$. The temperature is fixed at
$T=90$ MeV.} \label{fig:P-rhob-e}
\end{center}
\end{figure}

In Fig. \ref{fig:P-rhob-e}, we report the pressure
at $T=90$ MeV as a function of baryon density (in units of nuclear
saturation density $\rho_{0}=0.153$ fm$^{-3}$) (left panel) and
energy density (right panel). It is interesting to observe that
pressure as a function of baryon density (or energy density) is
stiffer in the pure hadronic phase for $q>1$ but appears a strong
softening in the mixed phase. This feature results in significant
changes in the incompressibility and may be particularly important
in identifying the presence of nonextensive effects in high energy
heavy ion collisions experiments. Related to this aspect, let us
observe that possible indirect indications of a significative
softening of the EOS at the energies reached at AGS have been
discussed several times in the literature \cite{prl}.

\begin{figure}
\begin{center}
\resizebox{0.6\textwidth}{!}{%
  \includegraphics{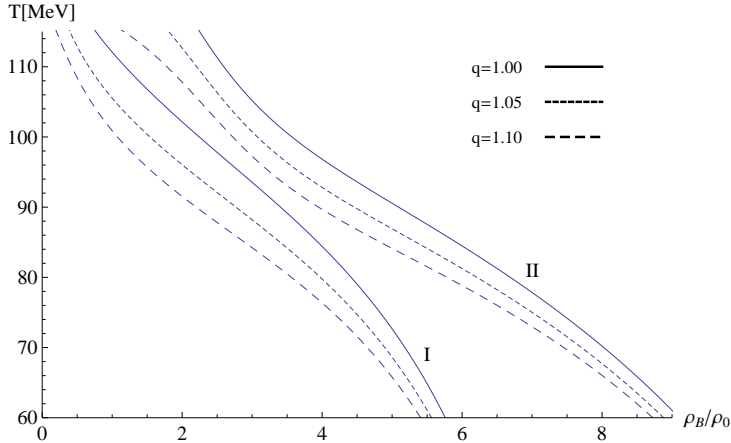} }
  \caption{Phase diagram $T-\rho_B$ for different values of $q$. The
curves with index $I$ and $II$ indicate, respectively, the
beginning and the end of the mixed phase.} \label{fig:RhoCT}
\end{center}
\end{figure}

In Fig. \ref{fig:RhoCT}, it is reported the phase diagram in the
plane $T-\rho_B$ for different values of $q$. The curves labelled
with the index $I$ and $II$ represent, respectively, the beginning
and the end of the mixed phase. For $q>1$, both the first and the
second critical densities are sensibly reduced, even if the shape
of the mixed phase is approximately the same. Related to this
aspect, let us mention that the simplest version of the MIT bag
model, considered in this investigation, appears to be not fully
appropriate to describe a large range of temperature and density.
To overcome this shortcoming, a phenomenological approach can
therefore be based on a density or temperature dependent bag
constant \cite{prl,burgio}. Moreover,
in regime of high temperature and small baryon
chemical potential the first order phase transition may end in a
(second order) critical endpoint with a smooth crossover. These
features cannot be incorporated in the considered mean field
approach. In our investigation, because we are focusing to
nonextensive statistical effects on the nuclear EOS, instead of
introducing additional parameterizations, we work with a fixed bag
constant and limit our analysis to a restricted range of
temperature and density, region of particular interest for high
energy compressed nuclear matter experiments.

\begin{figure}
\begin{center}
\resizebox{0.6\textwidth}{!}{%
  \includegraphics{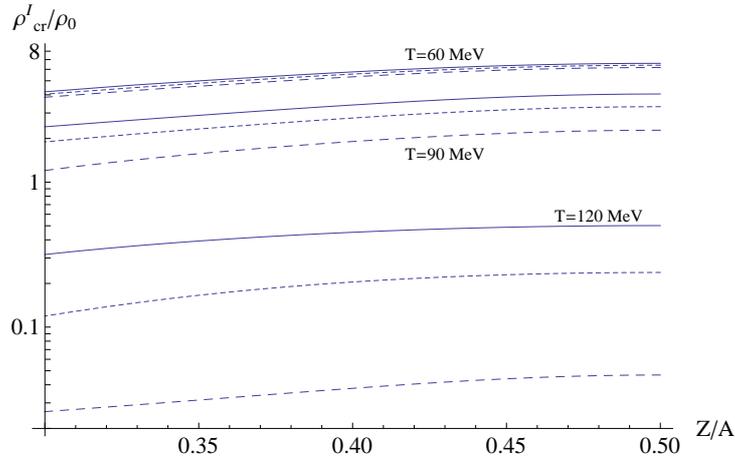} }
  \caption{Variation of the first transition baryon
  density as a function of the net electric charge fraction $Z/A$ for different temperatures and
  values of $q$ ($q=1$, solid lines; $q=1.05$, short dashed lines;
$q=1.10$, long dashed lines).}
  \label{fig:RhoCZA}
\end{center}
\end{figure}

Let us now explore in more details the variation of the first
transition baryon density $\rho^{I}_{\rm cr}$ as a function of
different physical parameters. In Fig. \ref{fig:RhoCZA}, we report
the dependence of $\rho^{I}_{\rm cr}$ as a function of $Z/A$ for
different values of $q$ ($y$ axis in logarithmic scale). It is
interesting to note a significant reduction of $\rho^{I}_{\rm cr}$
in presence of nonextensive statistics; as in the previous cases,
this effect increases with the temperature.
The dependence of the first transition baryon density as a
function of $Z/A$ is essentially a consequence of the $\rho$ meson
field behavior in the hadronic phase because it is directly
connected with the isospin density of the system (as appears from
Eq.(\ref{eq:MFT})). In this context, let us observe that, at fixed
value of $q$, $\rho^{I}_{\rm cr}$ is significantly reduced by
decreasing $Z/A$ only at lower temperatures ($T=60$ MeV) while, as
expected, at higher temperatures ($T=120$ MeV) the transition
baryon density becomes very low and its isospin dependence becomes
negligible, also in the framework of nonextensive statistics. This
matter of fact is a consequence of fact that at low baryon chemical potentials (or baryon
densities) the $\rho$ meson field becomes almost constant and its
absolute value significantly decreases.

\section{Conclusions}\label{conclusion}

Following the basic prescriptions of the Tsallis' nonlinear
relativistic thermodynamics, we investigate the relevance of
nonextensive statistical effects on the relativistic nuclear and
subnuclear equation of state. We have focused our investigation in regime
of finite temperature and baryon chemical potential, reachable in
high-energy heavy-ion collisions, for which the deconfinement
phase transition can be still considered of the first order.

In the first part of the work, we have investigated the hadronic
equation of state and the role played by the meson fields in the
framework of a relativistic mean field model which contains the
basic prescriptions of nonextensive (nonlinear) statistical mechanics. We have
shown that, also in presence of small deviations from standard
Boltzmann-Gibbs statistics, the meson fields and, consequently,
the EOS appear to be sensibly modified. In the second part, we
have analyzed the QGP proprieties using the MIT Bag model and also
in this case the EOS becomes stiffer in presence of nonextensive
effects. Finally, we have studied the proprieties of the phase
transition from hadronic matter to QGP and the formation of a
relative mixed phase by requiring the Gibbs conditions on the
global conservation of baryon number and electric charge fraction.
We have seen that nonextensive effects play a crucial role in the
deconfinement phase transition. Moreover, although pressure as a
function of baryon density is stiffer in the hadronic phase, we
have shown that a strong softening in the mixed phase takes place
in presence of nonextensive statistics. Such a behavior implies an
abruptly variation in the incompressibility and could be
considered as a signal of nonextensive statistical effects in high
energy heavy ion collisions.

From a phenomenological point of view, the nonextensive index $q$ is
considered here as a free parameter, even if, actually should not
be treated as such because, in principle, it should depend on the
physical conditions generated in the reaction, on the fluctuation
of the temperature and be related to microscopic quantities (such
as, for example, the mean interparticle interaction length, the
screening length and the collision frequency into the parton
plasma). Moreover, let us remember that, in the diffusional approximation,
a value $q>1$ implies the presence of a superdiffusion among the
constituent particles (the mean square displacement obeys to a
power law behavior $\langle x^2\rangle\propto t^\alpha$, with
$\alpha>1$) \cite{tsamem}.

\end{document}